\documentclass[aps,twocolumn,a4paper,amsmath,amssymb]{revtex4}

\usepackage{amssymb}
\usepackage{enumerate}
\usepackage{graphicx}
\usepackage{hyperref}
\usepackage[a4paper,left=2cm,right=2cm,top=2cm,bottom=2cm]{geometry}

\begin{document}

\title{How long does it take a society to learn a new term?}

\author{Sadegh Raeisi}
  \email{s.raeisi@physics.sharif.ir}

\affiliation{Department of Physics, Sharif University of
Technology,\\P.O. Box 11365-9161, Tehran, Iran}

\begin{abstract}
In this paper, I study the diffusion of new terms, called neologism, in
social networks. I consider it as an example of information
dynamics on networks and I hope that solving this problem can
help us to understand and describe the information dynamics
problem. To do so I develop a phenomenological model for the
diffusion mechanism. I find an analytical relationship between
number of people in the society who has learned the term and time
taken.  The Network parameters are imported in this analytical
solution. I also present some simulation for this mechanism for
several sample and some real networks which confirms the analytical results. In
addition, I study the effects of network topology on diffusion
process.
\end{abstract}
\maketitle

\section{introduction}
Recently, attempts to describe and understand the status and
different behaviors of a society have been the subject of intense
research. The structure of social networks has been studied and
simulated by different types of random graphs such as scale free
and small world graphs\cite{randomgraphs}. Some methods were
developed to identify communities in social networks
\cite{community1},\cite{community2},\cite{community3},\cite{community4},
and centrality in networks\cite{centrality}. There is no need to
talk about its importance because it is easy to realize the
effect of such an understanding on our life.

Besides these attempts to study the societies, some new researches
were developed to study phenomena that occur in the societies.
One of the most important phenomena in the society is that when a
piece of information enters the society, some people get it and
then it starts to spread through out the network. The piece of
information may be a gossip, a virus, fame of a film or a person
and etc. This phenomenon is called information spread or
information dynamics on networks.

The information spread has been studied in several cases. For
instance, the problem of gossip spread in social networks has
been studied by Lind {\em et al} \cite{gossip}. Similar researches were devoted to the
dynamic of reputation \cite{reputation} and the diffusion of
viruses \cite{viruses}, disease \cite{disease} and computer
viruses \cite{computerviruses}.

In this work, I go through the spread of a new term in a society.
Indeed, the main question I want answer is that, when a new
scientific term or expression is produced, how does it spread
through the scientific societies, but in this paper, I have
studied and solved the general problem of the spread of any term
in any social network.

To do this, first I used graphs as a mathematical tool to
describe the society in my model. Second, I present a
phenomenological model to illustrate the diffusion mechanism
where by phenomenological I mean that this model is mostly based
on our intuitional understanding of this phenomenon. In other
words, in this model, as we may expect, the term is generated by
someone out of the society and spreads through people's
conversations.

Here, I derive an analytical relationship between the number of
people who have learnt the term and time taken. One of the key
point in this job is that this relationship matches our
expectations. In addition, I developed some simulations for this
model which confirm the analytical results.

The structure of this paper is as follows: in section (\ref{model}), I
introduce my problem and my model. In section (\ref{analytical}) I present my
analytical solution for term spread in society. In section (\ref{simulation}) I
explain some details about the simulation that I developed for
this problem. Then some of the key results are extracted in
section (\ref{result}). Finally, some concluding remarks are made in Section (\ref{discussion}).

\section{The model}
\label{model}
Here I describe my model and explain the main concepts,
assumptions and my notation.

\begin{figure*}  \label{game}
 \centering
	 \includegraphics[width=13.0cm]{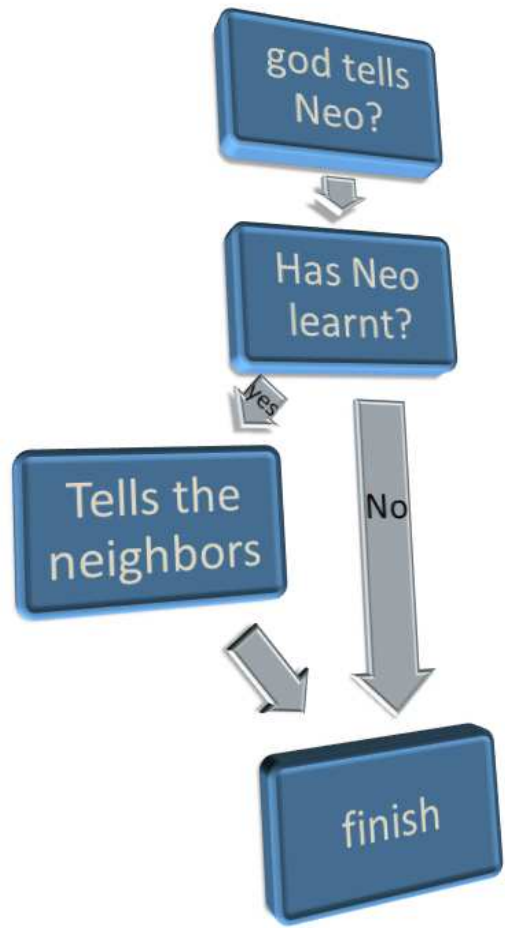}
  \caption{ step1) god chooses Neo randomly step2) if Neo learns the term, he teaches it to
    his friends and the game is finished, otherwise the game is
    finished}
\end{figure*}

I model the society as a graph G(N,E) where vertices N represent
people and edges E represent their relationships. In other words
a society is a set of vertices which are connected to each other
if and only if corresponding persons are related to each other.
This relation might be family relationship, friendship, a
scientific cooperation, etc. Here, I use simple graphs for
simplicity; directed and weighted graphs can be used in
appropriate situations. For example if people's relationships are
not the same, I can use weighted graphs.

On the other hand, I have to model the spread mechanism. The main
idea in my model  is that someone learns the new term and then
teaches it to those who he or she has a relation with. In this
section I explain the implementation of this idea.

To do this, I first need to implement the individual learning
process because the society learning is based on it. As in a real
situation, each person needs to hear the term several times to
learn it. Indeed, every person hears the term time after time
during the learning process, but he cannot use it unless the
number of times he has heard the term, \textbf{$d_{i}$}, passes a
specific number of times which I call the learning threshold
\textbf{$D_{i}$} and depends on his abilities. The threshold,
\textbf{$D_{i}$} is one of the model's parameters.

I also divide the society learning process into smaller
processes. Indeed my model is based on repeating a learning
step which I call game Fig. (~\ref{game}). In each game, someone is chosen
randomly and is called Neo. Neo hears the term and $d_{Neo}$
increases by one. If he learns the term, he would tell it to his
friends. So his friends hear the term once. Otherwise, if Neo does
not learn the term, he cannot use it in his conversations. After
Neo's action the game finishes.

In each game, some people hear the term and their $d$ increases
by one and after several games their hearing times reaches the
threshold $D$ and they learn the term. In other words, in each
game some people may learn the term also it may happen that no
new person learns it. Repeating the game, the number of learnt
people increases and approaches the total number of people in the
society, N.

In brief, the main characteristics of game are :
\begin{description}
\item[god]the one out side the game who knows the term and chooses the Neo and tells the term to him.
\item[Neo] the one who is randomly chosen in each game by god.
\item[learner] the one who has not learnt the term yet and $d<D$ for him.
\item[learnt]the one who has learnt the term and who can teach it to his friends.
\end{description}

Also, there are some assumptions in my model which are as follows
:
\begin{enumerate}
    \item The society network is static. In fact I assumed
    that people and their relationship are fixed when they learn
    term. Because the time scale of society dynamics seems to be negligible in comparison with learning process time scale, my assumption is still valid.
    \item There is a god out side the society who knows the term
    and chooses Neo and as an external force, tells him the term. In a
    real society, different choices such as multimedia, language
    academies, etc can play it's role.
    \item The leaning process for all terms are the same.
\end{enumerate}

Having defined the model, I explain my approach to use this model
to illustrate the learning process.

I first need to quantify the learning process and present a
measure for it. The number of learnt people seems to be a good
quantity for this purpose. Alternatively, I need a measure to
quantify the time in the game and a good choice for time is the
number of games. In each society, each game takes some time to
finish and the average of this time may be calculated easily. Consequently
each game represents a time scale. Now the number of learnt people
in each game, represent the learning improvement up to that time.
Using these two, I can easily define other needed quantities such
as learning speed. So, my model is complete.

From now on, for better and easier understanding, I fix the
following notations:
\begin{itemize}
    \item \textbf{g} \,\:Number of finished games
    \item \textbf{w} \,\:Number of learnt people
    \item \textbf{N} \,\:Number of people in the society
    \item \textbf{L} \,\:Number of relationship between people($\mid E \mid $) 
    \item \textbf{D} \,\:Average personal learning threshold
    \item \textbf{$d_{i}$} \,\:Number of times that the i th person has heard the term.
    \item \textbf{$k_{i}$} \,\:Number of people who are related to i
    th person in the network or the degree of i th vertex
    \item \textbf{$\overline{k}$} \,\:Average number of friend each person may have
\end{itemize}

I have studied this model with two approaches: first,
analytical solution of learning improvement W(g). Second,
simulation of model and studying the behavior of w versus g. Next
section focuses on analytical solution.

\section{analytical solution}
\label{analytical}

In this section, I want to derive an analytical relation between
learning improvement w and spent time g. One of the key point in
my job is that I make use of mean field approximation to do this.
This means that the distribution of quantities over a randomness
is represented by their average value.

Here, I first calculate the learning improvement per game
$\frac{\Delta w}{ \Delta g}$, then I assume it to be
$\frac{dw}{dg}$ which is valid when the learning steps are increased
in number and decreased in length. Consequently a differential
equation is obtained and solving the differential equation would
result in an analytical solution of W(g).

To calculate the learning improvement per game I first estimate
the number of people who hear the term in a game, $n_{game}$ and
then I can obtain the improvement in learning. To estimate the
average number of people called in a game, $n_{game}$, I follow
the game steps. In each game at least one person, Neo, is called.
Then depending on NEO's state, different cases are possible. Neo
may be in:
\begin{enumerate}
    \item Standby state:Neo has not learnt the term yet and does not learn in this
    game either. So he can not tell the term to his friends.
    \item Transition state: Neo has just heard the term D-1 times so he would pass his learning threshold in this game
    and would be able to tell the term to his friends from now on.
    \item Active state: NEO has learnt the term in previous games, so his knowledge does not improves,
    but he can tell the term to his friends.
\end{enumerate}
Number of learnt persons differs in each case. So I calculate the
average number of people who learn the term in each game. To do so
I need to calculate the probability of occurrence of each
case ${p_{standby}, p_{transition}, p_{active}}$ and the number of
people who learn the term in each case ${n_{stand
by}, n_{transition}, n_{active}}$. Then the average number of people
who learn term in a game, $n_{game}$ can be calculate as :

\begin{eqnarray} \label{wgame}
  \overline{n} _{game} &=& p_{standby}*n_{standby} \cr
		&+& p_{transition}*n_{transition}+p_{active}* n_{active}
\end{eqnarray}

So I need just to calculate the {$p_{i}$} and {$n_{i}$} sets where i
is a possible case.

On the first step I calculate the {$p_{i}$} set. I can make use of
the total number of learnt people, $W$ to guess the Neo's state.
$\frac{W}{N}$ indicates the probability of a person being learnt
and therefore the probability of Neo being in active state. On
the other hand, $(1-\frac{W}{N})$ denotes the probability of Neo
being in standby or transition state. As each person needs to
hear the term $D$ times to learn it, on first$(D-1)$ times he would
not pass the threshold and remains in standby state, but in $D$ th
time which is the transition state he joins the learnt persons.
Thus, the chance of being in standby state is $(D-1)$'th times more
than transition state. As a result the probability of occurrence
of these states are obtained as:
\begin{eqnarray}\label{s}
    P_{standby}&=&(1-\frac{W}{N})(\frac{D-1}{D}) \cr
    P_{transition}&=&(1-\frac{W}{N})(\frac{1}{D})\cr
    P_{active}&=& \frac{W}{N}
\end{eqnarray}

The next step is calculating the {$n_{i}$} set and for Neo's
different states, they are easily obtained as :
\begin{description}
    \item[Standby:] Only his own knowledge improves : $n_{standby}=1$
    \item[Transition:] As his knowledge improves he becomes able to call his
    friends:$n_{transition}=1+\overline{k}$
    \item[Active:] He can call his friends but his knowledge does not
    improves anymore: $n_{active}=\overline{k}$
\end{description}
Now according to Eq.(~\ref{wgame}) I can calculate $\overline{n}_{game}$
\begin{equation}\label{w}
    \overline{n}_{game}:=\sum_{i=1}^{3}{n_{i}P_{i}=A+\frac{W}{N}(\overline{k}-A)}
\end{equation}
where
\begin{equation}\label{A}
    A:=\frac{d+ \overline{k}}{D}
\end{equation}
There is just one step left to calculate the learning improvement
per game. The point is that among all those who have heard the
term during the game, only the knowledge of those who have not
learned the term, increases. In other words, if those who have
heard the term are in standby or transition states their
knowledge increases, so I have to multiply $\overline{n}_{game}$
by $P_{stand by}+P_{transition}$ to obtain the average number of
people whose knowledge improves. To determine the amount of
improvement per each game, it suffices to determine the amount of
improvement when each person is called. As explained person$_{i}$
needs to hear the term $D$th times, so each time that he is
called, his knowledge improves by $\frac{1}{D}$. Finally the following differential
equation is obtained:
\begin{equation}\label{ana-dif}
    \frac{dW}{dg}=\frac{1}{D}(A+\frac{W}{N}(\overline{k}-A))(1-\frac{W}{N})
\end{equation}
again with:
\begin{equation}\label{A}
    A:=\frac{D+ \overline{k}}{D}
\end{equation}
Solving the differential equation leads to:
\begin{equation}\label{W(g)}
    W(g)=N\frac{e^{-\frac{\overline{k}*g}{N*D}}-1}{(\frac{A-\overline{k}}{A})e^{-\frac{\overline{k}*g}{N*D}}-A}
\end{equation}

This equation ends the analytical solution. It is just left to
check how good this relation describes the society learning. To do
so, I developed and ran some simulation which is illustrated in
the next section.

\section{simulation}
\label{simulation}

In this section I want to explain my simulation procedure, some
details about it and some of its features.

First, I need to implement the society. To do so, I make use of
a mathematical graph. Indeed, I produce a $M_{N*N}$ matrix which
is the adjacency matrix of the graph. Each row and column in this
matrix represents a person in the society and person a is a friend
of person b if $M_{a,b}=1$.

To construct the adjacency matrices, I used the real network data of condensed matter collaborations collected by Mark Newman\cite{realdata}. In addition, I made use of pajeck and ORA softwares to generate some random graphs such as scale free and small world
and tested my model on generated graphs.

Then, for each person i, I set a $D_{i}$ randomly between 1 and 100
which means that people's learning threshold is between 1 and 100.

Second, I use the adjacency matrix to implement the game. Each
time a person is chosen randomly as Neo and his learning $d_{Neo}$
increases by one. Then if $d_{Neo}\geq D_{Neo}$, according to
adjacency matrix Neo's friends are recognized and corresponding
$d$s increases. Now the game is finished and a new game would
start.

I also defined an other parameter, w which increases each time
that someone's learning passes his threshold. This parameter
indicates the number of learnt people in the society.

The game is repeated until a certain percent of society,
$\alpha=W/N$ learn the term and the number of games is counted.
Also, in this program, I calculate the number of past games for each
$\alpha$, 100 times and average over all the results. Consequently
the number of games which is required to have a certain percentage
of the society learn the term, is obtained.

\begin{figure*}\label{overview}
 \centering
 \includegraphics[width=7.0 cm]{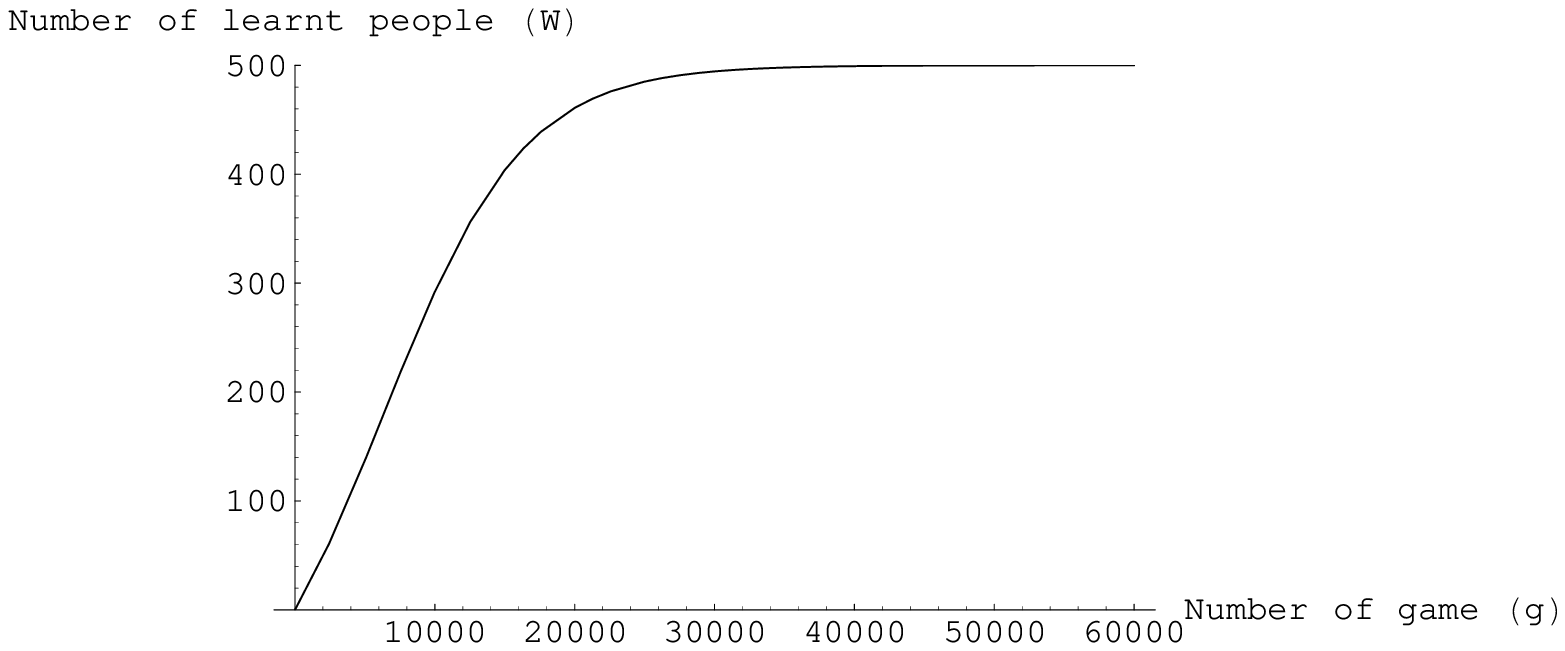}\hfill
 \includegraphics[width=6.0 cm]{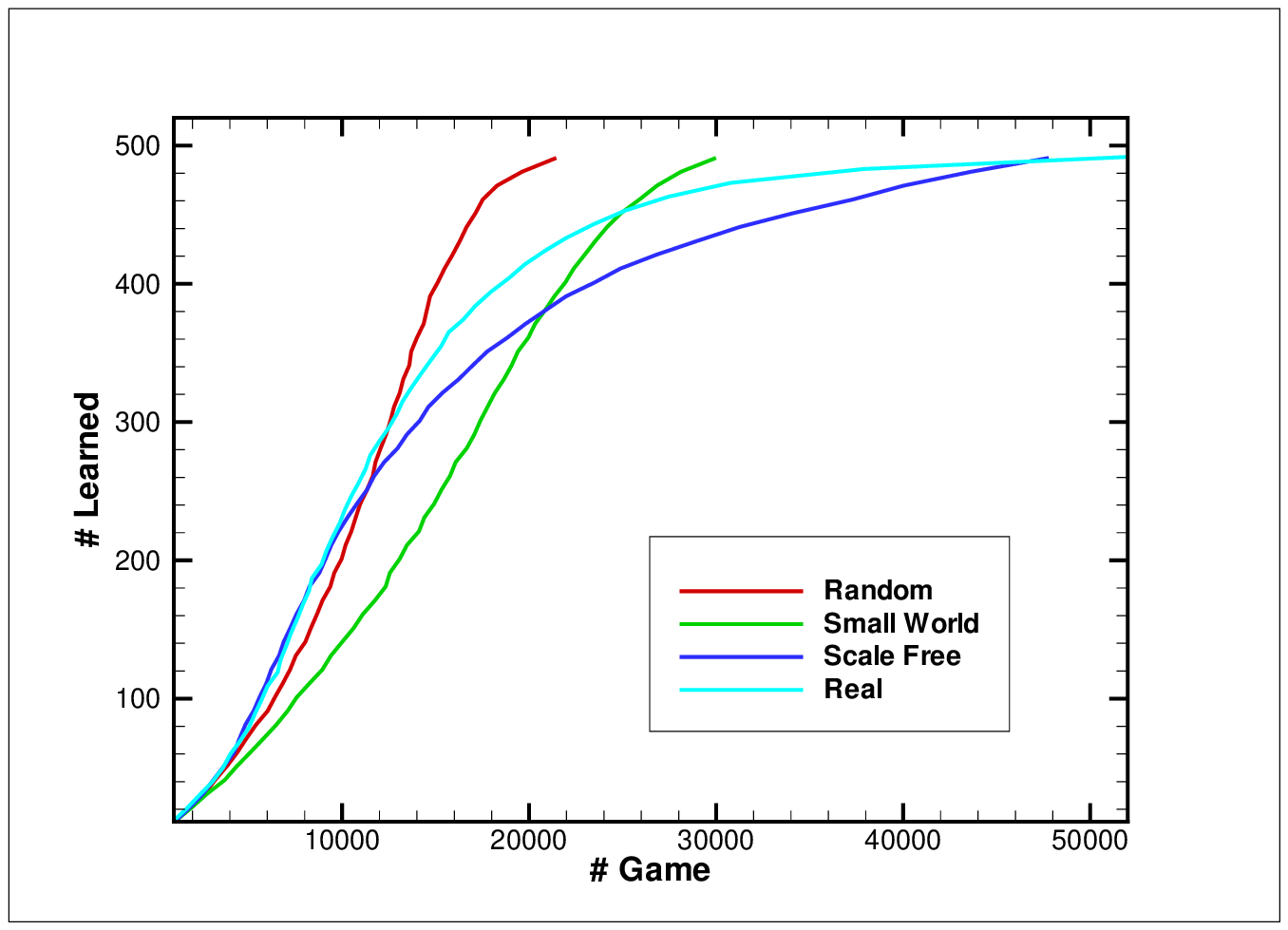}\hfill\null
\caption{a)learning behavior according to analytical solution for a network with N=500, and L=1500.
b)learning behavior according to simulation result for different
network topologies with N=500 and L=1500} 
\end{figure*}

I repeat this procedure for $\alpha = 0, .01, .02, ..., .98$ to
obtain the relation between $\alpha$ and number of games. Fig. (~\ref{overview} -b)
presents my results for several networks, one real, one random,
one scale free and one small world network.

I used and analyzed the simulation results. Next section go
through a comparison between simulation results and analytical
solution.
\section{results}
\label{result}
In this section I first study the overall behavior of the
mechanism according to both analytical solution and simulation
results and then its dependence on model parameters such as
number of edges and of vertices. I also check the results to be
logical and match our expects.
\subsection{learning process}
The analytical solution describes the learning process as follows
: at the beginning, the learning velocity is slow and only few
people learn the term during first games. Then this velocity
would speed up gradually till a specific percentage of the
society learns the term and then it slows down again.
Consequently, only few people learn during the last games. Fig. (~\ref{overview})
presents the learning process according to both analytical
solution and simulations for different network topologies. It is
seen that simulation results confirm the analytical results for
the learning process.

\begin{figure}\label{velocity}
 \centering
	 \includegraphics[width=6.0cm]{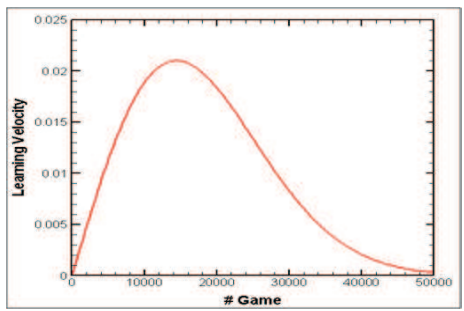}
  \caption{the speed of learning process versus number of past games for a network with N=500 and L=1500.}
\end{figure}

In addition, Fig. (~\ref{velocity}) shows the learning velocity of the society
versus time (g). It starts with zero, then speeds up to a maximum,
and then slows down again for large values of g or large time and
limits to zero.

Next two sections illustrate the effect of network topology on
learning process.
\subsection{learning process and the size of population(N)}
Here I go through the effect of size of the population, N. To do
so I work with the percent of society who have learned the term
and games taken:
\begin{figure} \label{overN}
 \centering
 	\includegraphics[width=6.0cm]{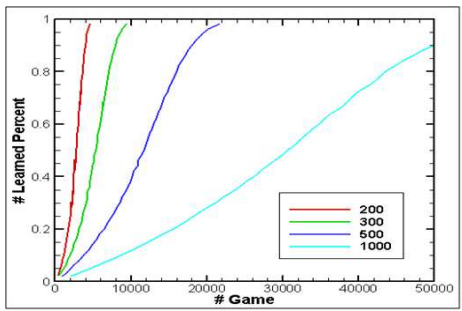}
 \caption{Learning process for networks of different sizes, N and fixed distribution function and relationship size of L=1500}
\end{figure}

\begin{equation}\label{saturation}
    W(g)=\alpha N
\end{equation}
I can calculate number of games taken for the
desired percent according to Eq. (~\ref{W(g)}). Solving Eq. (~\ref{saturation}), I can find the game in which $\alpha$
percent of the society learns the term. I denote it as $g^{*}$
and it is obtained as:
\begin{equation}\label{gstar}
    g^{*}=\frac{D*N Log[\frac{A(\alpha -1)-\overline{k}\alpha}{A(\alpha -1)}]}{\overline{k}}
\end{equation}
From now on, I work with $g^{*}(\alpha)$ as the saturation time
for a desired amount of $\alpha=\frac{W}{N}$.

As a result of both analytical solution and simulation results,
societies with larger population need more time to be saturated which seems to be natural.
Fig. (~\ref{overN}) presents the simulation results for learning
process for societies of different sizes. It is seen
that for larger N, the learning process graph shifts to larger
time and saturates slower which means that for larger population, larger
saturation time, $g^{*}$ is needed.

On the other hand, according to Eq. (~\ref{gstar}), the saturation time
depends linearly on population size, N which means that analytical
solution confirms the simulation results.

\subsection{learning process and the number of relationship(L)}
\begin{figure} \label{overL}
 \centering
 	\includegraphics[width=6.0cm]{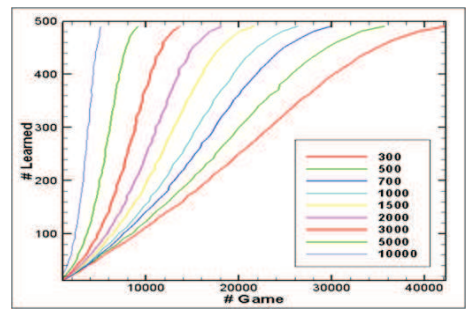}
 \caption{Learning process for networks with different number of edges, L and fixed distribution function and N=500} 
\end{figure}

The other result is about the effect of people relationship, L on
the learning process. Imagine a society in which most people know
each other versus a society in which only few people are
connected. In this section I describe how learning process
differs for these two societies. Again both simulation and
analytical results are presented. Fig. (~\ref{overL}) shows learning process
for networks with different number of edges. These
figures, indicates that as people relationships increase the
society learns the term sooner and the time scale of saturation
decreases.

\begin{figure}\label{overL-a}
 \centering
 	\includegraphics[width=6.0cm]{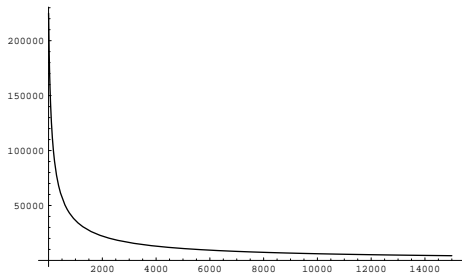}
 \caption{analytical prediction for saturation time versus L for N=500}
\end{figure}

On the other hand, the analytical solution predicts similar
behaviors. Obviously, the $\overline{k}$ is proportional to number
of edges L :
\begin{equation}\label{kbar}
    \overline{k}=\frac{s L}{N}
\end{equation}
Where s is a constant that depends on the distribution function of
degrees of vertices of the network and for a fixed distribution
function it does not change. Replacing Eq. (\ref{kbar}) in Eq. (~\ref{gstar}), the
following relation is obtained:
\begin{equation}
    g^{*} = {\frac{d N^2 Log[\frac{L s (-1+\alpha
    )+d (N (-1+\alpha )-L s \alpha )}{(d N+L s)
    (-1+\alpha )}]}{L s}}
\end{equation}
This equation is plotted in Fig. (~\ref{overL-a}) and describes the analytical
behavior of saturation time $g^{*}$ versus L. According to this
analytical result, when the size of relation between people, L
increases, society learns the term faster and in fact the
saturation time decreases which is compatible with both
simulation result and our expectations.
\section{conclusion}
\label{discussion}

I have studied the process in which a new term spreads out
through a social network.  I have used a phenomenological model to
describe this process and studied this model both analytically
and using computer simulation.

According to this job, the process in which a new term spreads
out through a network is described as follows: At the beginning of
the process, few people knows the term. Consequently, the role of
people interaction in learning process is negligible in compassion
with external force. But as time passes, the number of learnt
people increases  and their contribution in learning process
increases the process velocity. It means that the number of
people who learn the term in a specific time increases. Finally
when most of the society learn the term, as there are few learner
people, the  process velocity  decreases and limits to zero. This
description is illustrated in Fig. (~\ref{overview}).

Maybe the key feature of my job is that I have imported the
topological parameter of the society network in my model. In
other word, network topology is accounted in my job. For instance
it were seen that when the number of people, N increases, the
saturation time increases in turn or when the number of
edge increases in network, the saturation time
decreases.

For the next step the relation between different terms in a language may be considered. For instance the interrealtion between terms may affect the society learning proccess and help people to learn it faster. Also the spread mechanism can be improved or other kind of diffusion mechanism maybe checked.
\section{acknowledgment }
\label{acknowledgment}
I want to thank F. Ghanbarnejad first because of her nice consults and second because she helped me through the simulation. Indeed, she implemented the spread mechanism. I also want to acknowledge Prof. Mansouri and S. Hamidi for their discussions.
\bibliographystyle{amsplain}
\bibliography{gh}

\begin{thebibliography}{99}
\bibitem{randomgraphs}M. E. J. Newman, S. H. Strogatz, and D. J.Watts, Phys. Rev. E
\textbf{64}, 026118 (2001).

\bibitem{community1}D. Wilkinson and B. A. Huberman, Proc. Natl. Acad. Sci. USA, Feb. 2, 2004,
10.1073/pnas.0307740100.

\bibitem{community2}L. Freeman, Sociometry, vol. 40, pp. 35–41, 1977.

\bibitem{community3}U. Brandes, Journal of Mathematical Sociology, vol. 25, no. 2, pp. 163–177,
2001.

\bibitem{community4}M. E. J. Newman, cond-mat/0309508/, 2003. F. Wu and B. A.

\bibitem{community4}Huberman, Europhysics Letters, 2004.

\bibitem{centrality} L. C. Freeman, Social Networks, 1(1978/79)
215-239.

\bibitem{gossip}Pedro G. Lind,  Luciano R. da Silva, Jos´e S. Andrade Jr., and Hans J. Herrmann, Phys. Rev. E \textbf{76},
036117 (2007).

\bibitem{reputation}B. A. Huberman, F. Wu, Journal of Statistical Mechanics:
Theory and Experiment(2004)P04006.

\bibitem{viruses}Z. Dezso and A.-L. Barabasi,  Phys. Rev. E, \textbf{65}, 055103 (2002).

\bibitem{disease}M. Newman, Phys. Rev. E, \textbf{66}, 016128 (2002).

\bibitem{computerviruses}M. E. J. Newman, S. F., and J. Balthrop, Phys. Rev. E, \textbf{66}, 035101 (2002).

\bibitem{realdata} M. E. J. Newman, Proc. Natl. Acad. Sci. USA \textbf{98}, 404-409 (2001).

\end{thebibliography}
{}

\end{document}